\documentclass[11pt,a4paper]{article}
\usepackage[utf8]{inputenc}
\usepackage[T1]{fontenc}
\usepackage{amsmath,amssymb}
\usepackage{graphicx}
\usepackage[margin=1in]{geometry}
\usepackage{booktabs}
\usepackage{hyperref}
\usepackage[numbers]{natbib}
\usepackage{xcolor}
\usepackage{float}
\usepackage{caption}

\hypersetup{
    colorlinks=true,
    linkcolor=blue,
    citecolor=blue,
    urlcolor=blue
}

\title{\textbf{``All You Need'' is Not All You Need for a Paper Title:\\On the Origins of a Scientific Meme}}

\author{
  Anton Alyakin\\
  \texttt{alyakin314@gmail.com}
}

\date{}

\begin{document}

\maketitle

\begin{abstract}
The 2017 paper ``Attention Is All You Need'' introduced the Transformer architecture—and inadvertently spawned one of machine learning's most persistent naming conventions. We analyze 717 arXiv preprints containing ``All You Need'' in their titles (2009–2025), finding exponential growth (R² > 0.994) following the original paper, with 200 titles in 2025 alone. Among papers following the canonical ``X [Is] All You Need'' structure, ``Attention'' remains the most frequently claimed necessity (28 occurrences). Situating this phenomenon within memetic theory, we argue the pattern's success reflects competitive pressures in scientific communication that increasingly favor memorability over precision. Whether this trend represents harmless academic whimsy or symptomatic sensationalism, we leave—with appropriate self-awareness—to the reader.
\end{abstract}

\section{Introduction}

In June 2017, \citet{vaswani2017attention} published a paper that would 
fundamentally reshape the landscape of deep learning. While on the surface 
appearing to be yet just another machine translation paper, ``Attention Is 
All You Need'' introduced the  Transformer architecture which was destined 
to become the backbone of modern  large language models~\citep{brown2020gpt},
vision-language  models~\citep{radford2021learning}, and countless other
applications.
However, just as serendipitous as its technical impact was its cultural legacy:
a title format that has proven irresistible to subsequent researchers.

The phrase ``All You Need Is Love,'' first broadcast to an estimated 400 million
viewers during The Beatles' 1967 appearance on \textit{Our
World}~\citep{beatles1967love}, carries an inherent memetic appeal. Its
structure---asserting that a single element is sufficient for some complex
all-capturing endeavor---combines simplicity with boldness. When applied to
academic work, this format implicitly promises parsimony: the paper will
identify one crucial insight that renders other approaches unnecessary. While
this indeed happened with the attention mechanism replacing nearly all recurrent
and convolutional components in sequence modeling, it has since been used
hundreds of times, often with considerably less revolutionary justifications,
implications, and consequences.

This study examines the proliferation of the ``All You Need'' naming pattern in
academic manuscripts indexed on arXiv. We address three questions: 
\begin{itemize}
    \item How did the frequency of this pattern evolve over time?
    \item What subjects do researchers claim are ``all you need''?
    \item What can this phenomenon tell us about academic culture and the dynamics of scientific communication?
\end{itemize}

\section{Methodology}

\subsection{Data Collection}

We queried the arXiv API for all papers containing the phrase ``all you need''
in their titles. The search was conducted on December 2nd, 2025 and returned
717 papers spanning 2009--2025. For each paper, we extracted the title,
publication date, and the arXiv category.

\subsection{Pattern Analysis}

To analyze the structure of titles, we identified the words preceding
the verb in the canonical ``X [Is/Are/Is Not/etc.] All You Need'' format. 
This yielded 499 matching papers (69.6\% of the corpus). We experimented 
with extracting subjects following the phrase (i.e., ``All You Need Is X'')
but found only 21 entries with no repeated subjects.

\section{Results}

\subsection{The Pre-Transformer and Early Transformer Eras}

Table~\ref{tab:early} presents all arXiv preprints with ``All You Need'' in
their titles published before 2019. This chronological listing reveals the
pattern's sparse early usage and subsequent explosion following the Transformer
paper. 

Notably, the first instance of ``All You Need'' appeared in high energy physics
(HEP)~\citep{cherman2009baryon}, predating the machine learning adoption by six
years. The second~\citep{mishkin2016init} and third~\citep{xie2017beyond}
preprints address neural network initialization, notably with the latter
explicitly referencing the former. The fourth paper is the ``Attention
Is All You Need'' itself~\citep{vaswani2017attention}.
The fifth manuscript, named, ironically, ``CNN Is All You 
Need''~\citep{chen2017cnn} is a machine translation paper and appears to be the
first work with a name intentionally referencing \citet{vaswani2017attention}.
Interestingly, note that \citet{grondahl2018love} appear to be referencing the 
Beatles, despite being published after \citet{vaswani2017attention}.

\begin{table}[htbp]
\centering
\caption{Complete list of ``All You Need'' arXiv preprints published before 2019.}
\label{tab:early}
\small
\begin{tabular}{cp{11.5cm}cc}
\toprule
\textbf{\#} & \textbf{Title} & \textbf{Year} & \textbf{Citation} \\
\midrule
1 & All you need is N: Baryon spectroscopy in two large N limits & 2009 & \cite{cherman2009baryon} \\
2 & All you need is a good init & 2015 & \citep{mishkin2016init} \\
3 & All You Need is Beyond a Good Init: Exploring Better Solution for Training Extremely Deep Convolutional Neural Networks with Orthonormality and Modulation & 2017 & \citep{xie2017beyond} \\
4 & Attention Is All You Need & 2017 & \citep{vaswani2017attention} \\
5 & CNN Is All You Need & 2017 & \citep{chen2017cnn} \\
6 & Diversity is All You Need: Learning Skills without a Reward Function & 2018 & \citep{eysenbach2018diversity} \\
7 & MemGEN: Memory is All You Need & 2018 & \citep{gelly2018memgen} \\
8 & Cross-lingual Argumentation Mining: Machine Translation (and a bit of Projection) is All You Need! & 2018 & \citep{eger2018crosslingual} \\
9 & All You Need is "Love": Evading Hate Speech Detection & 2018 & \citep{grondahl2018love} \\
10 & Bytes are All You Need: End-to-End Multilingual Speech Recognition and Synthesis with Bytes & 2018 & \citep{li2018bytes} \\
\bottomrule
\end{tabular}
\end{table}

\pagebreak

\begin{figure}[t]
\centering
\includegraphics[width=\textwidth]{figure1_frequency_over_time.png}
\caption{Left: Annual publication counts showing accelerating growth post-2017. Right: Cumulative papers with exponential fit.}
\label{fig:growth}
\end{figure}

\subsection{Exponential Temporal Dynamics - Catching Fire}

Figure~\ref{fig:growth} presents the annual distribution and cumulative growth 
of ``All You Need'' papers. The publication of \citet{vaswani2017attention} 
marked an inflection point, after which growth became approximately exponential. 
The year 2025 alone accounts for 200 papers (27.9\% of all papers collected), 
with the trend showing no signs of saturation at the time of data collection. 
Fitting an exponential model to the cumulative data yields an excellent fit 
($R^2 > 0.994$), suggesting the meme's spread follows dynamics similar to 
other viral phenomena.

\subsection{Spread to Other Categories}

\begin{figure}[htbp]
\centering
\includegraphics[width=\textwidth]{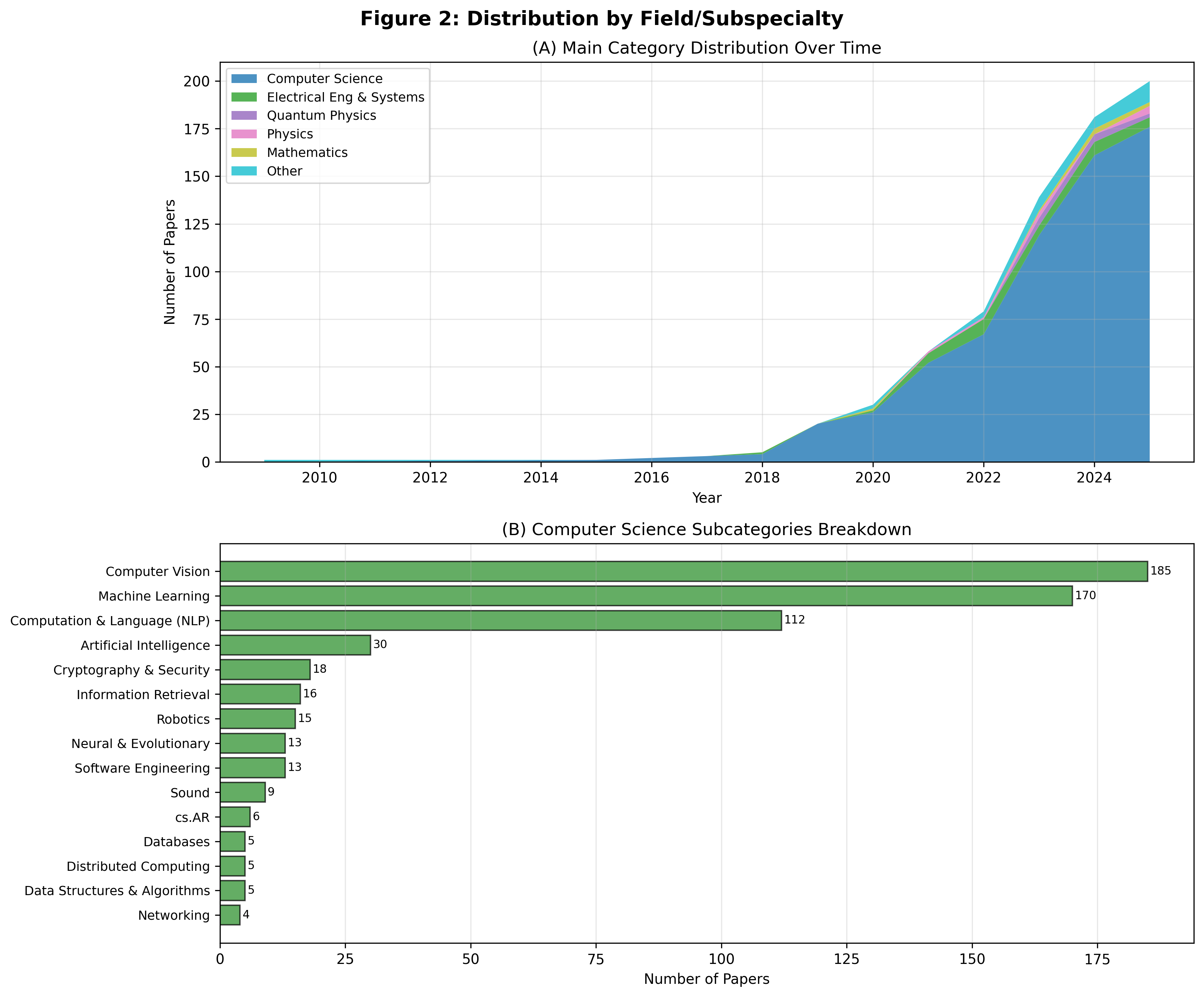}
\caption{Top: Category distribution over time. Bottom: Breakdown of Computer Science subcategories.}
\label{fig:category_growth}
\end{figure}

As shown in Figure~\ref{fig:category_growth} and the main category distribution in 
Table~\ref{tab:main_cat}, the proliferation of the ``All You Need'' meme is overwhelmingly 
concentrated in computer science (87.7\% of all titles). The subfield 
breakdown in Table~\ref{tab:detailed_cat} expectedly demonstrates the dominance of artificial 
intelligence-adjacent areas with Computer Vision, Machine Learning, and NLP 
collectively accounting for the majority of usages. 

Despite this strong centralization, the pattern has clearly begun to diffuse
outward. Table~\ref{tab:first_appear}, which lists first appearances by category, shows that over
time the naming convention has emerged in areas as diverse as quantitative 
biology, condensed matter physics, astrophysics, and finance. While these 
incursions remain sparse compared to computer science, their presence 
demonstrates that the meme has crossed disciplinary boundaries and is now 
present in fields far removed from deep learning.

\begin{table}[htbp]
\centering
\begin{minipage}{0.38\textwidth}
\centering
\captionof{table}{Main category distribution.}
\label{tab:main_cat}
\small
\begin{tabular}{l r}
\toprule
\textbf{Main Category} & \textbf{Count (\%)} \\
\midrule
Computer Science & 629 (87.7\%) \\
Electrical Eng \& Systems & 32 (4.5\%) \\
Quantum Physics & 10 (1.4\%) \\
Physics & 9 (1.3\%) \\
Mathematics & 7 (1.0\%) \\
Statistics & 6 (0.8\%) \\
Condensed Matter & 6 (0.8\%) \\
HEP - Phenomenology & 5 (0.7\%) \\
Quantitative Biology & 5 (0.7\%) \\
Astrophysics & 3 (0.4\%) \\
HEP - Lattice & 2 (0.3\%) \\
Nonlinear Sciences & 1 (0.1\%) \\
HEP - Experiment & 1 (0.1\%) \\
Quantitative Finance & 1 (0.1\%) \\
\bottomrule
\end{tabular}
\end{minipage}%
\hfill
\begin{minipage}{0.58\textwidth}
\centering
\captionof{table}{Top 20 detailed categories.}
\label{tab:detailed_cat}
\small
\begin{tabular}{l l r}
\toprule
\textbf{Code} & \textbf{Category} & \textbf{Count (\%)} \\
\midrule
cs.CV & Computer Vision & 185 (25.8\%) \\
cs.LG & Machine Learning & 170 (23.7\%) \\
cs.CL & Computation \& Language & 112 (15.6\%) \\
cs.AI & Artificial Intelligence & 30 (4.2\%) \\
cs.CR & Cryptography \& Security & 18 (2.5\%) \\
cs.IR & Information Retrieval & 16 (2.2\%) \\
eess.IV & Electrical Eng \& Systems & 15 (2.1\%) \\
cs.RO & Robotics & 15 (2.1\%) \\
eess.AS & Electrical Eng \& Systems & 13 (1.8\%) \\
cs.NE & Neural \& Evolutionary & 13 (1.8\%) \\
cs.SE & Software Engineering & 13 (1.8\%) \\
quant-ph & Quantum Physics & 10 (1.4\%) \\
cs.SD & Sound & 9 (1.3\%) \\
cs.AR & Computer Science & 6 (0.8\%) \\
hep-ph & HEP - Phenomenology & 5 (0.7\%) \\
cs.DB & Databases & 5 (0.7\%) \\
stat.ML & Statistics & 5 (0.7\%) \\
cs.DC & Distributed Computing & 5 (0.7\%) \\
cs.DS & Data Structures \& Algorithms & 5 (0.7\%) \\
cs.NI & Networking & 4 (0.6\%) \\
\bottomrule
\end{tabular}
\end{minipage}
\end{table}

\begin{table}[htbp]
\centering
\caption{First appearance of ``All You Need'' papers by category.}
\label{tab:first_appear}
\small
\begin{tabular}{r l p{9.5cm}}
\toprule
\textbf{Year} & \textbf{Category} & \textbf{Paper Title} \\
\midrule
2009 & HEP - Phenomenology & All you need is N: Baryon spectroscopy in two large N limits \citep{cherman2009baryon} \\
2015 & Computer Science & All you need is a good init \citep{mishkin2016init} \\
2018 & Electrical Eng \& Systems & Bytes are All You Need: End-to-End Multilingual Speech Recognition and Synthesis with Bytes \citep{li2018bytes} \\
2020 & Mathematics & Scalability and robustness of spectral embedding: landmark diffusion is all you need \citep{shen2020scalability} \\
2020 & Statistics & All You Need is a Good Functional Prior for Bayesian Deep Learning \citep{tran2020functional} \\
2020 & Quantitative Biology & Evolution Is All You Need: Phylogenetic Augmentation for Contrastive Learning \citep{lu2020evolution} \\
2021 & Physics & All you need is time to generalise the Goman-Khrabrov dynamic stall model \citep{ayancik2021time} \\
2022 & Nonlinear Sciences & Physics-Enhanced Bifurcation Optimisers: All You Need Is a Canonical Complex Network \citep{syed2022bifurcation} \\
2022 & Condensed Matter & Microscopy is All You Need \citep{kalinin2022microscopy} \\
2023 & Quantum Physics & All you need is spin: SU(2) equivariant variational quantum circuits based on spin networks \citep{east2023spin} \\
2023 & HEP - Lattice & Equivariant Transformer is all you need \citep{tomiya2023equivariant} \\
2023 & HEP - Experiment & Two Watts is All You Need: Enabling In-Detector Real-Time Machine Learning for Neutrino Telescopes Via Edge Computing \citep{jin2023twowatts} \\
2024 & Astrophysics & Five parameters are all you need (in $\Lambda$CDM) \citep{monterocamacho2024five} \\
2024 & Quantitative Finance & MoA is All You Need: Building LLM Research Team using Mixture of Agents \citep{chen2024moa} \\
\bottomrule
\end{tabular}
\end{table}

\subsection{Semantic Distribution}

\begin{figure}[htbp]
\centering
\includegraphics[width=0.65\textwidth]{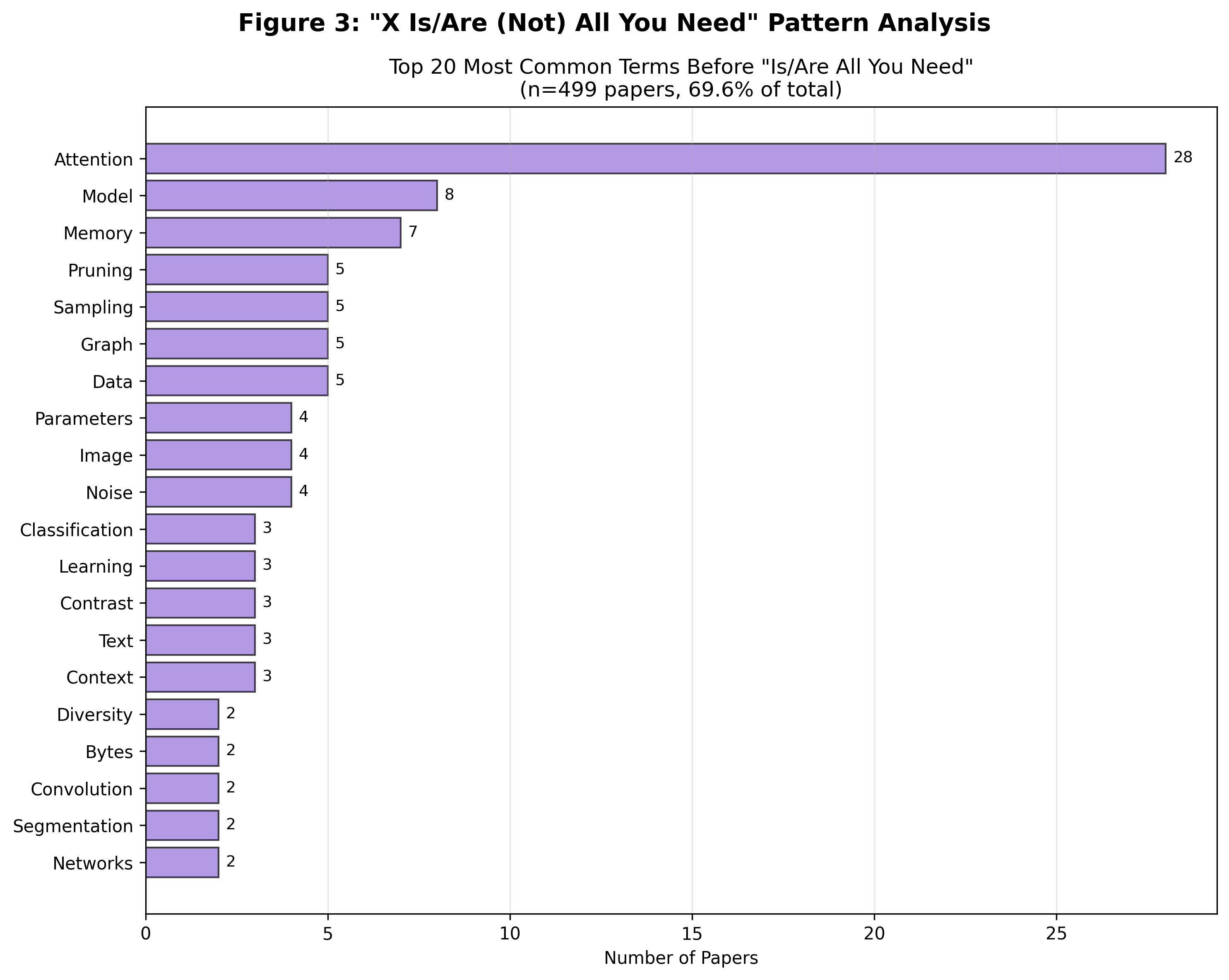}
\caption{Most common single-word subjects in the ``X [Is] All You Need'' pattern.}
\label{fig:semantic}
\end{figure}

Figure~\ref{fig:semantic} presents the most frequently occurring words in the 
``X [Is] All You Need'' pattern. Unsurprisingly, ``Attention'' leads with 28 
occurrences---a testament both to the original paper's influence and to 
subsequent researchers' attempts to refine or challenge its claims.

\pagebreak

\section{Discussion}

\subsection{The Anatomy of an Academic Meme}

The concept of the meme, as originally articulated by
\citet{dawkins1976selfish}, describes a unit of cultural transmission that
propagates through imitation, analogous to genes in biological evolution.
Dawkins defined memes as "tunes, ideas, catch-phrases, clothes fashions, ways of
making pots or of building arches" that replicate by "leaping from brain to
brain." 

The ``All You Need'' pattern exhibits the three characteristics Dawkins
identified as essential for successful replicators: longevity (the pattern has
persisted for over eight years), fecundity (717 instances and counting), and
copying fidelity (the core structure remains remarkably stable across
variations). By adopting the format, authors align their work with a lineage of
influential research while simultaneously claiming a degree of boldness. The
structure's flexibility---any noun or phrase can precede ``Is All You 
Need''---makes it applicable to nearly any contribution across most domains.

\subsection{Comparison with Other Scientific Naming Conventions}

The ``All You Need'' pattern is not the first formulaic titling convention to
achieve widespread adoption in academic literature. Mathematics and physics
have long favored the ``On X'' construction---exemplified by Turing's
foundational ``On Computable Numbers''~\citep{turing1936computable} and
Einstein's ``On the Electrodynamics of Moving  
Bodies''~\citep{einstein1905electrodynamics}---a convention dating to classical 
treatises, perhaps signaling scholarly modesty. The machine learning literature 
has developed its own ecosystem of such patterns; for example ``Towards X'' 
signals incremental progress with a clear ambitious end-goal in 
mind~\citep{ren2016towards}.

What distinguishes the ``All You Need'' pattern is its implicit promise of 
parsimony---the claim that complexity can be reduced to a single sufficient 
element. While ``Towards'' hedges with epistemic humility and ``On'' maintains 
disciplinary neutrality, ``All You Need'' asserts completeness. This boldness 
may explain both its appeal and its ironic overuse: in a competitive attention 
economy, modesty is a disadvantage.

\subsection{Implications for Scientific Communication}

The exponential growth of ``All You Need'' pattern occurs against a backdrop of
intensifying competition for attention in academic publishing. The ``publish or
perish'' culture---a phrase with roots extending back to at least the late
1920s~\citep{case1928scholarship, cabanac2018primordial}---has created
structural incentives that reward visibility alongside---or sometimes instead
of---substance. Publications attract funding, and funding enables further
publication, creating feedback loops that favor attention-grabbing presentations
of research. Visibility increasingly influences success. This is especially true
in the modern world, and in the field, where much of the dissemination of science 
bypasses the formal academic peer-review and happens via self-submitted 
preprint sharing.\footnote{In addition to granting the work credibility by 
publishing it, the publishing venues also serve as a de-sensationalization filter, imposing uniform standards. Preprint and published titles often vary 
dramatically---e.g., the author's ``Metformin: We need to either put it 
in our drinking water or rethink how we study it''~\citep{alyakin2024metformin_preprint} 
became ``Exploration of Residual Confounding in Analyses of Associations 
of Metformin Use and Outcomes in Adults With Type 2 Diabetes''~\citep{alyakin2024metformin_published} 
upon publication.}

The pattern's success may also reflect the increasing role of social media in
scientific discourse. Memorable titles serve legitimate functions: they aid
recall and facilitate discussion. Preprints shared on Twitter (now X) or Reddit
benefit from memorable titles that can be easily quoted and recognized. In this
environment, ``Attention Is All You Need'' functions not merely as a title but
as a slogan---a compression of the paper's contribution into a form optimized
for viral transmission.

\subsection{Limitations}

Our analysis is limited to arXiv and thus excludes papers published only in
traditional venues. We intentionally omitted author level analysis.

\subsection{A Self-Referential Note}

The author acknowledges the irony of contributing yet another ``All You Need''
paper to the literature, albeit in negated form. \citet{hofstadter1979geb}
devoted considerable attention to such ``strange loops''---self-referential
structures in which, by moving through a hierarchical system, one unexpectedly
returns to one's starting point. By analyzing the ``All You Need'' phenomenon with
a title that participates in it (albeit does not include it due to temporal
causation), this paper necessarily adds itself to the dataset it describes,
achieving a modest Hofstadterian recursion.

\section{Conclusion}

We have presented an empirical analysis of the ``All You Need'' naming pattern
in machine learning and related fields. Our findings document exponential growth
from 3 papers pre-2017 to 717 by late 2025. The pattern is showing no signs of
saturation with 200 papers already in 2025 alone. ``Attention'' remains the most
claimed necessity, appearing in 28 titles. The conclusion of whether the
persistence of this naming convention represents a harmless quirk of academic
culture or a symptom of declining originality and rising 
sensationalism-to-subject ratio is left up to the readers and future researchers.
Perhaps, they
will use more original titles than this work, since, in the end, good science
with an appropriate but memorable title is all one needs. 

\section*{Remark}

This work is intentionally written to be light-hearted but thought-provoking.
It is in part inspired by Pretraining on the Test Set Is All You
Need~\citep{schaeffer2023pretrainingtestsetneed}.
The author deeply respects the researchers conducting work in the field of
artificial intelligence, and appreciates a good reference and the use of humor. There has been a number of breakthrough works with the names following the
``All You Need'' convention, especially in the field of tiny models.

\section*{Reproducibility}

The code and date is available at \url{https://github.com/alyakin314/All_You_Need}.

\section*{Acknowledgments}

The code and manuscript of this work were drafted with the help of 
Anthropic Claude Opus 4.5.

\bibliographystyle{unsrtnat}
\bibliography{references}

@inproceedings{vaswani2017attention,
  title={Attention is All You Need},
  author={Vaswani, Ashish and Shazeer, Noam and Parmar, Niki and Uszkoreit, Jakob and Jones, Llion and Gomez, Aidan N and Kaiser, {\L}ukasz and Polosukhin, Illia},
  booktitle={Advances in Neural Information Processing Systems},
  volume={30},
  pages={5998--6008},
  year={2017}
}

@article{cherman2009baryon,
  title={All you need is {N}: Baryon spectroscopy in two large {N} limits},
  author={Cherman, Aleksey and Cohen, Thomas D and Lebed, Richard F},
  journal={arXiv preprint arXiv:0906.2400},
  year={2009}
}

@inproceedings{mishkin2016init,
  title={All You Need is a Good Init},
  author={Mishkin, Dmytro and Matas, Jiri},
  booktitle={International Conference on Learning Representations},
  year={2016}
}

@article{xie2017beyond,
  title={All You Need is Beyond a Good Init: Exploring Better Solution for Training Extremely Deep Convolutional Neural Networks with Orthonormality and Modulation},
  author={Xie, Di and Xiong, Jiang and Pu, Shiliang},
  journal={arXiv preprint arXiv:1703.01827},
  year={2017}
}

@article{chen2017cnn,
  title={{CNN} Is All You Need},
  author={Chen, Qiming and Wu, Ren},
  journal={arXiv preprint arXiv:1712.09662},
  year={2017}
}

@article{eysenbach2018diversity,
  title={Diversity is All You Need: Learning Skills without a Reward Function},
  author={Eysenbach, Benjamin and Gupta, Abhishek and Ibarz, Julian and Levine, Sergey},
  journal={arXiv preprint arXiv:1802.06070},
  year={2018}
}

@article{gelly2018memgen,
  title={{MemGEN}: Memory is All You Need},
  author={Gelly, Sylvain and Kurach, Karol and Michalski, Marcin and Szepesvari, Csaba},
  journal={arXiv preprint arXiv:1803.11203},
  year={2018}
}

@inproceedings{eger2018crosslingual,
  title={Cross-lingual Argumentation Mining: Machine Translation (and a bit of Projection) is All You Need!},
  author={Eger, Steffen and Daxenberger, Johannes and Stab, Christian and Gurevych, Iryna},
  booktitle={Proceedings of the 27th International Conference on Computational Linguistics},
  pages={831--844},
  year={2018}
}

@inproceedings{grondahl2018love,
  title={All You Need is ``Love'': Evading Hate-speech Detection},
  author={Gr{\"o}ndahl, Tommi and Pajola, Luca and Juuti, Mika and Conti, Mauro and Asokan, N},
  booktitle={Proceedings of the 11th ACM Workshop on Artificial Intelligence and Security},
  pages={2--12},
  year={2018}
}

@article{li2018bytes,
  title={Bytes are All You Need: End-to-End Multilingual Speech Recognition and Synthesis with Bytes},
  author={Li, Bo and Zhang, Yu and Sainath, Tara and Wu, Yonghui and Chan, William},
  journal={arXiv preprint arXiv:1811.09021},
  year={2018}
}

@misc{beatles1967love,
  title={All You Need Is Love},
  author={{The Beatles}},
  year={1967},
  note={Single released July 7, 1967. Parlophone Records}
}

@article{turing1936computable,
  title={On Computable Numbers, with an Application to the {Entscheidungsproblem}},
  author={Turing, Alan M.},
  journal={Proceedings of the London Mathematical Society},
  volume={s2-42},
  number={1},
  pages={230--265},
  year={1936}
}

@book{hofstadter1979geb,
  title={G{\"o}del, Escher, Bach: An Eternal Golden Braid},
  author={Hofstadter, Douglas R.},
  year={1979},
  publisher={Basic Books},
  address={New York}
}

@article{case1928scholarship,
  author = {Case, Clarence Marsh},
  title = {Scholarship in Sociology},
  journal = {Sociology and Social Research},
  volume = {12},
  pages = {323--340},
  year = {1927--1928}
}

@article{cabanac2018primordial,
  author = {Cabanac, Guillaume},
  title = {What is the primordial reference for ...?---Redux},
  journal = {Scientometrics},
  volume = {114},
  number = {2},
  pages = {481--488},
  year = {2018},
  doi = {10.1007/s11192-017-2595-4}
}

@misc{schaeffer2023pretrainingtestsetneed,
      title={Pretraining on the Test Set Is All You Need}, 
      author={Rylan Schaeffer},
      year={2023},
      eprint={2309.08632},
      archivePrefix={arXiv},
      primaryClass={cs.CL},
      url={https://arxiv.org/abs/2309.08632}, 
}

@article{einstein1905electrodynamics,
  title={Zur {Elektrodynamik} bewegter {K{\"o}rper}},
  author={Einstein, Albert},
  journal={Annalen der Physik},
  volume={322},
  number={10},
  pages={891--921},
  year={1905},
  note={English translation: ``On the Electrodynamics of Moving Bodies''}
}

@book{dawkins1976selfish,
  title={The Selfish Gene},
  author={Dawkins, Richard},
  year={1976},
  publisher={Oxford University Press},
  address={Oxford}
}

@misc{alyakin2024metformin_preprint,
  author    = {Powell, Mike and Clark, Callahan and Alyakin, Anton and Vogelstein, Joshua T. and Hart, Brian},
  title     = {Metformin: We need to either put it in our drinking water or rethink how we study it},
  journal   = {medRxiv},
  year      = {2021},
  doi       = {10.1101/2021.09.15.21263634},
  url       = {https://www.medrxiv.org/content/10.1101/2021.09.15.21263634v1},
  note      = {Preprint}
}

@article{alyakin2024metformin_published,
  author    = {Powell, Mike and Clark, Callahan and Alyakin, Anton and Vogelstein, Joshua T. and Hart, Brian},
  title     = {Exploration of Residual Confounding in Analyses of Associations of Metformin Use and Outcomes in Adults With Type 2 Diabetes},
  journal   = {JAMA Network Open},
  year      = {2022},
  volume    = {5},
  number    = {11},
  pages     = {e2241505},
  doi       = {10.1001/jamanetworkopen.2022.41505},
  url       = {https://doi.org/10.1001/jamanetworkopen.2022.41505},
  month     = nov
}

@article{shen2020scalability,
  title={Scalability and robustness of spectral embedding: landmark diffusion is all you need},
  author={Shen, Chao and Wu, Hau-Tieng},
  journal={arXiv preprint arXiv:2001.00801},
  year={2020}
}

@article{tran2020functional,
  title={All You Need is a Good Functional Prior for Bayesian Deep Learning},
  author={Tran, Ba-Hien and Rossi, Simone and Milios, Dimitrios and Filippone, Maurizio},
  journal={arXiv preprint arXiv:2011.12829},
  year={2020}
}

@article{lu2020evolution,
  title={Evolution Is All You Need: Phylogenetic Augmentation for Contrastive Learning},
  author={Lu, Amy X and Lu, Amy X and Moses, Alan},
  journal={arXiv preprint arXiv:2012.13475},
  year={2020}
}

@article{ayancik2021time,
  title={All you need is time to generalise the Goman-Khrabrov dynamic stall model},
  author={Ayancik, Fatma and Mulleners, Karen},
  journal={arXiv preprint arXiv:2110.08516},
  year={2021}
}

@article{syed2022bifurcation,
  title={Physics-Enhanced Bifurcation Optimisers: All You Need Is a Canonical Complex Network},
  author={Syed, Marvin and Berloff, Natalia G},
  journal={arXiv preprint arXiv:2207.11256},
  year={2022}
}

@article{kalinin2022microscopy,
  title={Microscopy is All You Need},
  author={Kalinin, Sergei V and others},
  journal={arXiv preprint arXiv:2210.06526},
  year={2022}
}

@article{east2023spin,
  title={All you need is spin: SU(2) equivariant variational quantum circuits based on spin networks},
  author={East, Richard D P and Alonso-Linaje, Guillermo and Park, Chae-Yeun},
  journal={arXiv preprint arXiv:2309.07250},
  year={2023}
}

@article{tomiya2023equivariant,
  title={Equivariant Transformer is all you need},
  author={Tomiya, Akio and Nagai, Yuki},
  journal={arXiv preprint arXiv:2310.13222},
  year={2023}
}

@article{jin2023twowatts,
  title={Two Watts is All You Need: Enabling In-Detector Real-Time Machine Learning for Neutrino Telescopes Via Edge Computing},
  author={Jin, Miaochen and Hu, Yushi and Arg{\"u}elles, Carlos A},
  journal={arXiv preprint arXiv:2311.04983},
  year={2023}
}

@article{monterocamacho2024five,
  title={Five parameters are all you need (in $\Lambda$CDM)},
  author={Montero-Camacho, Paulo and others},
  journal={arXiv preprint arXiv:2405.13680},
  year={2024}
}

@article{chen2024moa,
  title={MoA is All You Need: Building LLM Research Team using Mixture of Agents},
  author={Chen, Sandy and Zeng, Leqi and Raghunathan, Abhinav and Huang, Flora and Kim, Terrence C},
  journal={arXiv preprint arXiv:2409.07487},
  year={2024}
}

@inproceedings{brown2020gpt,
  author    = {Brown, Tom and Mann, Benjamin and Ryder, Nick and Subbiah, Melanie and Kaplan, Jared D and Dhariwal, Prafulla and Neelakantan, Arvind and Shyam, Pranav and Sastry, Girish and Askell, Amanda and Agarwal, Sandhini and Herbert-Voss, Ariel and Krueger, Gretchen and Henighan, Tom and Child, Rewon and Ramesh, Aditya and Ziegler, Daniel and Wu, Jeffrey and Winter, Clemens and Hesse, Chris and Chen, Mark and Sigler, Eric and Litwin, Mateusz and Gray, Scott and Chess, Benjamin and Clark, Jack and Berner, Christopher and McCandlish, Sam and Radford, Alec and Sutskever, Ilya and Amodei, Dario},
  booktitle = {Advances in Neural Information Processing Systems},
  title     = {Language Models are Few-Shot Learners},
  volume    = {33},
  pages     = {1877--1901},
  year      = {2020}
}

@inproceedings{radford2021learning,
  title     = {Learning Transferable Visual Models From Natural Language Supervision},
  author    = {Radford, Alec and Kim, Jong Wook and Hallacy, Chris and Ramesh, Aditya and Goh, Gabriel and Agarwal, Sandhini and Sastry, Girish and Askell, Amanda and Mishkin, Pamela and Clark, Jack and Krueger, Gretchen and Sutskever, Ilya},
  booktitle = {Proceedings of the 38th International Conference on Machine Learning},
  pages     = {8748--8763},
  year      = {2021},
  volume    = {139},
  series    = {Proceedings of Machine Learning Research},
  publisher = {PMLR}
}

@misc{ren2016towards,
      title={Faster R-CNN: Towards Real-Time Object Detection with Region Proposal Networks}, 
      author={Shaoqing Ren and Kaiming He and Ross Girshick and Jian Sun},
      year={2016},
      eprint={1506.01497},
      archivePrefix={arXiv},
      primaryClass={cs.CV},
      url={https://arxiv.org/abs/1506.01497}, 
}

\end{document}